\documentclass[%
 reprint,
superscriptaddress,
nofootinbib,
 amsmath, amssymb,
 aps,
pra,
]{revtex4-2}

\usepackage{graphicx}
\graphicspath{ {/paper/}}
\usepackage{dcolumn}
\usepackage{bm}
\usepackage{slashed}
\usepackage{amsmath}
\usepackage{amssymb}
\usepackage{amsfonts}
\usepackage{enumitem}
\usepackage{blindtext} 
\usepackage[linktocpage]{hyperref}
\usepackage[svgnames]{xcolor}
\hypersetup{
     colorlinks = true,
     linkcolor = magenta,
     anchorcolor = black,
     citecolor = blue,
     filecolor = blue,
     urlcolor = blue
     }

\usepackage{bbold}
\usepackage{braket}
\usepackage{tikz}
\usetikzlibrary{quantikz2}
\usetikzlibrary{positioning}
\usepackage{kantlipsum}
\usetikzlibrary{shadows,arrows,calc,backgrounds,fit,shapes,snakes,shapes.multipart,decorations.pathreplacing,shapes.misc,patterns,positioning}
\usepackage{pgfplots}
\usetikzlibrary{backgrounds,fit,decorations.pathreplacing}

\begin{document}

\preprint{APS/123-QED}

\title{Dissipation-induced Quantum Homogenization for Temporal Information Processing}

\author{Alexander Yosifov}
\email{alexanderyyosifov@gmail.com}
\affiliation{Space Research and Technology Institute, Bulgarian Academy of Sciences, Sofia, 1113, Bulgaria}

\author{Aditya Iyer}
\email{aditya.iyer@physics.ox.ac.uk}
\affiliation{Clarendon Laboratory, University of Oxford, Parks Road, Oxford, OX1 3PU, UK}

\author{Vlatko Vedral}
\email{vlatko.vedral@physics.ox.ac.uk}
\affiliation{Clarendon Laboratory, University of Oxford, Parks Road, Oxford, OX1 3PU, UK}

\date{\today}

\begin{abstract}
Quantum reservoirs have great potential as they utilize the complex real-time dissipative dynamics of quantum systems for information processing and target time-series generation without precise control or fine-tuning of the Hamiltonian parameters. Nonetheless, their realization is challenging as quantum hardware with appropriate dynamics, robustness to noise, and ability to produce target steady states is required. To that end, we propose the disordered quantum homogenizer as an alternative interaction model, and prove it satisfies the necessary and sufficient conditions --- \textit{stability} and \textit{contractivity} --- of the reservoir dynamics, necessary for solving machine learning tasks with time-series input data streams. The results indicate that the quantum homogenization protocol, physically implementable as either nuclear magnetic resonance ensemble or a photonic system, can potentially function as a reservoir computer.  
\end{abstract}

\maketitle


\section{\label{sec:level1}Introduction}
Harnessing quantum systems for machine learning (ML) has emerged as a viable alternative, promising exponential computational advantages over existing ML models based on artificial neural networks or their neuromorphic-inspired physical versions \cite{1,qml}. In the current noisy intermediate-scale quantum (NISQ) era, quantum systems are particularly suitable for computation due to their exponentially scaling Hilbert spaces and ability to natively process quantum modes \cite{2, 2a}. Where devices of several dozen qubits have already performed notably faster in niche applications, demonstrably hard for digital computers, \textit{e.g.} integer factorization \cite{3, 3a}, and some combinatorial optimization problems \cite{4,5,optimization}, offering glimpses of the expected supremacy of quantum computers.

Beyond this narrow domain, however, the potential of real-time quantum information processing \cite{6, 6a} for practical ML problems, involving large constantly-evolving data streams, has not been realized. This is due to the fact that, typically, quantum systems are susceptible to noise, while some have inappropriate dynamics or need extensive control over their degrees of freedom \cite{review}. Evidently, any attempt to alleviate those challenges requires the utilization of quantum systems with non-trivial dynamics, satisfying certain necessary and sufficient conditions concerning their \textit{stability} and \textit{contractivity} \cite{7,8,9}, without the need for parameter fine-tuning.\footnote{The roles of stability and contractivity in classical reservoir computing were studied in \cite{lyudmila} (and more recently in \cite{simon,simon2}), while their relation to quantum reservoir computing was examined in \cite{ortega}.}

In this direction, the most promising framework is quantum reservoir computing (QRC) \cite{14} --- a class of recurrent neural networks (RNNs), where the \textit{reservoir} is a nonlinear processor of interacting qubits that leverages the real-time quantum dissipative dynamics as internal memory. In the standard QRC implementation, a low-dimensional input is projected into the higher-dimensional reservoir which facilitates both the encoding and the separation of inputs, similar to function mapping in kernel methods. So far, QRC has demonstrated its potential for many temporal ML problems \cite{10}, such as speech recognition \cite{11}, language processing \cite{12}, stock market movements \cite{13}, as well as universal quantum computation \cite{uni}, and supervised quantum ML \cite{qml2}. Besides, owing to recent advances in quantum hardware, it has even been physically realized via nuclear magnetic resonance (NMR) spin ensemble systems, with experimental demonstrations of \textit{learning} a nonlinear function with low mean square error \cite{15}, and state transformations \cite{17, 17a}.

Clearly, there are many possibilities for reservoir-based advancements in the realm of practical ML using available hardware. In light of this, we study the quantum homogenizer \cite{17, 17a} --- a quantum machine, experimentally realizable on a 4-qubit NMR system (or on photonic systems \cite{vlatko2}), composed of a reservoir of identical qubits, and utilized for quantum state approximation and steady state preparation \cite{16} via the partial $\mathtt{SWAP}$ --- unitary time-ordered bipartite interaction with time-varying coupling. More precisely, we analyze distance function inequalities for pairs of states, evolving under the isometric reservoir embedding of the homogenizer, and prove its dynamics is both \textit{stable} and \textit{contractive}.Therefore, by demonstrating satisfiability of the aforementioned necessary and sufficient conditions, our work introduces the homogenizer as a candidate interaction model for QRC with ML applications. 

\section{\label{sec:level2}Quantum homogenizer}
The homogenizer \cite{16, 17, 17a} is a quantum machine with dissipative dynamics, composed of a reservoir $\mathcal{R}$ of identical qubits $\{ \mathcal{R}_{k} \}_{k=1}^{N}$, all prepared in a canonical quantum state $\xi=\xi^{(k)}$ $\forall k \in N$.\footnote{The dissipative effects of the homogenizer are not unique. In fact, for a broader class of QRC and collision models \cite{uni,ieee}, engineered dissipation has been recognized as a resource for quantum computing, especially for quantum steady state preparation and entanglement stabilization schemes.} A single qubit $\hat{s}$ in arbitrary state $\rho_{\hat{s}}\in S(\mathcal{H}_{\hat{S}})$, injected into $\mathcal{R}$, sequentially interacts in a discrete-time manner $\{t_{i}\}\in\mathbb{Z}_{+}$ (where $\mathbb{Z_{+}} := \{0,1,2,\dots,\}$ and $\mathbb{Z}$ is the set of all integers) with its $N$ qubits via the unitary partial \texttt{$\mathtt{SWAP}$} operator $\tilde{\mathcal{U}}$

\begin{equation}
\label{eq:eq1}
\tilde{\mathcal{U}}_{N}^{\dagger}\dots\tilde{\mathcal{U}}_{1}^{\dagger}\left(\rho_{\hat{s}}\otimes \xi^{\otimes N}\right)\tilde{\mathcal{U}}_{1}\dots\tilde{\mathcal{U}}_{N}\approx\xi^{\otimes N+1}
\end{equation}
where $\rho^{(N)}_{\hat{s}} \rightarrow \xi$ as $N\rightarrow\infty$. Here, $\tilde{\mathcal{U}}_{k} := \tilde{\mathcal{U}}\otimes(\otimes_{j\neq k}\mathbb{1}_{j})$ is the bipartite interaction between the input qubit and the $k^{\text{th}}$ reservoir qubit, defined as\footnote{Although the analysis in the main text is of a single input qubit $\hat{s}$, the argument straightforwardly extends to arbitrarily long input sequences $\mathbf{\hat{s}}=\{{\hat{s}_{k}}\}^{N}_{k=1}$, where $\mathbf{\hat{s}}\in\{0,1\}$ or $\mathbf{\hat{s}}\in[0,1]$.}

\begin{equation}
\label{eq:unitaryoperation}
\tilde{\mathcal{U}} := \text{exp}\left(-\frac{i}{\hbar} \hat{s} \cdot \mathcal{R}_{k}\int J(t)dt\right)    
\end{equation}
where $J(t)\in\mathbb{R}$ is the time-varying coupling parameter, whose elements are drawn randomly each timestep from a uniform probability density $p(\psi_{1},\psi_{2})=p(U_{1}\psi_{2},U_{2}\psi_{2})$, and remain fixed during the interaction. The intensity of $\tilde{\mathcal{U}}$ depends on $J(t)$, where for $\int J(t)dt = \pi(\text{mod}\, 2\pi)$, $\tilde{\mathcal{U}}$ acts like the standard \texttt{$\mathtt{SWAP}$} gate, where \texttt{$\mathtt{SWAP}$}$\ket{\phi}\ket{\psi} = \ket{\psi}\ket{\phi}$, while for $J(t) \sim 0$, $\tilde{\mathcal{U}}\equiv\mathbb{1}$, where $\mathbb{1}: \ket{\psi}\rightarrow\ket{\psi}$. Intuitively, one can understand this protocol as a quantum information formalization of Markovian collision models \cite{collision2}, where thermalization is achieved via elementary interactions between an input system and a large bath of ancillas (for non-Markovianity, see \cite{collision1}).

The time variance of the coupling in Eq. (\ref{eq:unitaryoperation}), like in continuous-variable reservoirs (\textit{e.g.} oscillator-based models \cite{oscillator}), is crucial as it allows for more complex evolution and introduces nonlinearity to the dynamical encoding of the input into the higher-dimensional reservoir state space, while still being unitary, see \cite{7}. Recently, this was recognized to dramatically enhance the QRC performance, where its role in quantum ML was elucidated in \cite{nonlinear}. Here, we make use of the fact that in real-time computing encoding of time-varying input data (requiring memory) with fixed coupling, and fixed input with time-varying coupling each timestep (disordered dynamics) is quantum mechanically analogous and is known to lead to nonlinear evolution \cite{20}.

\begin{figure}[ht]
\centering
\includegraphics[width=\linewidth]{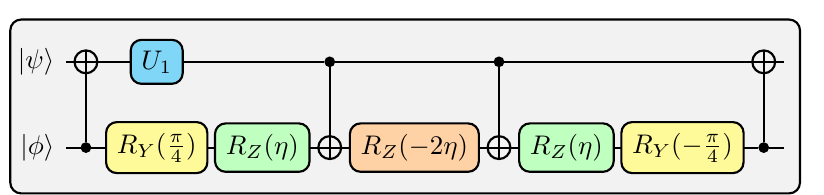}
\caption{Quantum circuit representation of the unitary transformation (\ref{eq:unitaryoperation}).}
\label{fig:figure}
\end{figure}

Meanwhile, for a $\mathcal{C}^{2}\otimes\mathcal{C}^{2}$ system, the quantum circuit, corresponding to the unitary transformation (\ref{eq:unitaryoperation}), can be natively realized on any current-era NISQ platform using four \texttt{CNOT} gates and six single-qubit gates, see Fig. \ref{fig:figure} \cite{16}.

Explicitly,

\begin{align}
U_{1} &= 
\begin{pmatrix} 
1 & 0 \\
0 & e^{\frac{i\pi}{2n}}
\end{pmatrix} \nonumber
\\
R_{Y} \left(\frac{\pi}{4}\right) &= \frac{1}{\sqrt{2}}
\begin{pmatrix}
1 & -1 \\
1 & 1
\end{pmatrix} \nonumber
\\
R_{Y} \left(-\frac{\pi}{4}\right) &= \frac{1}{\sqrt{2}}
\begin{pmatrix}
1 & 1 \\
-1 & 1
\end{pmatrix} \nonumber
\\
R_{Z} (\eta) &=
\begin{pmatrix}
e^{-i\eta} & 0 \\
0 & e^{i\eta}
\end{pmatrix} \nonumber
\\
R_{Z} (-2\eta) &=
\begin{pmatrix}
e^{2i\eta} & 0 \\
0 & e^{2i\eta}
\end{pmatrix} \nonumber
\end{align}
\\
where $-2\eta = \pi/n$. Physically, using (\ref{eq:unitaryoperation}) for computation can be advantageous (\textit{e.g.} solid state quantum computing \cite{ss1, ss2}) as only controlling the interaction time is necessary.

\section{\label{sec:level3}Stability and contractivity}
In principle, any quantum system (including quantum simulators and quantum annealers), exhibiting sufficiently rich dissipative dynamics, can be harnessed for processing time-series input data for practical ML problems \cite{7, 26}. Such diversity, and the ease of only using the natural system dynamics, compared to elaborate quantum gate-based alternatives, can lead to significant hardware simplifications in reservoir-based models. This facilitates the engineering of quantum systems based on simple bipartite input-reservoir unitary interactions, such that the associated dynamics is \textit{asymptotically stable} and \textit{contractive}, making them efficient information processors.\footnote{Recently, \cite{nak} demonstrated an extension of the echo state property to \textit{non-stationary} QRC models, illuminating important architecture advancements. Particularly, the finite variance output signals relative to initial-state difference decay, encountered in the noisy regime of the quantum homogenizer. While \cite{nak2}, similar to our work (and \cite{16}), highlighted the important role of environment interactions for the information retention capacity of reservoir systems.} In this work, we study the dynamics of the homogenizer and prove it satisfies those necessary and sufficient conditions, where we use the traditional treatment of the echo state property. In particular, we adopt the following definitions:\\

\textbf{Definition \hypertarget{definition}{1} (Stability).} A homogenizer \cite{16, 17, 17a} composed of $N$ qubits with reservoir dynamics (\ref{eq:eq1}), described by a bipartite unitary operator of the form (\ref{eq:unitaryoperation}) and a corresponding map (\ref{eq:cptp}) is asymptotically stable iff for any real-bounded input sequence ${\hat{S}}=\{{\hat{s}_{k}}\}^{N}_{k=1}$ (where $\text{sup}_{k\in\mathbb{Z}}|\hat{s}_{k}|<\infty$) in some initial logical state $\rho_{\hat{s}}^{(0)}\in S(\mathcal{H}_{\hat{S}})$ and for a reservoir in arbitrary canonical quantum state $\xi=\xi^{(k)}$ $\forall k \in N$, we have $\tilde{\mathcal{U}}_{N}:\rho^{(N)}_{\hat{s}}\rightarrow\xi$, such that $\lim_{N\rightarrow\infty} \parallel\prod^{N}_{k=1}\mathcal{M}^{(k)}\left(\rho_{\hat{s}}^{(0)}\right)\parallel_{2} = \xi$, where $\prod$ is the time-ordered composition of the family of input-independent completely positive trace preserving (CPTP) maps $\{\mathcal{M}^{(k)}\}$.\\

This sufficient condition can be understood as the quantum counterpart of the classical nonlinear input-output map approximation theorem in which sense the system has \textit{approximately finite memory} \cite{18,19}. So, if satisfied, the condition ensures that input $\hat{s}$ in any state $\rho_{\hat{s}}\in S(\mathcal{H}_{\hat{S}})$, injected into the higher-dimensional reservoir $\mathcal{R}$ in state $\xi\in S(\mathcal{H}_{\mathcal{R}})$, asymptotically converges towards $\xi$ under $\mathcal{M}$.\footnote{As we remark in the following Section, an essential feature of the homogenizer dynamics that we make use of is that it naturally converges arbitrary logical states towards the steady state $\xi^{(0)}$, which can be arbitrarily initialized.} Evidently, for the homogenizer to be asymptotically stable, the CPTP map $\mathcal{M}$ must be contractive. Hence, the reservoir stability guarantees the computations are performed independently of $\rho_{{\hat{s}}}^{(0)}$. That is to say, the evolution is solely determined by the unitary input-reservoir interactions and the effects of the initial logical state fade away as it gets smeared across the mutual correlations between the reservoir degrees of freedom via the repetitive application of (\ref{eq:cptp}).\\ 

\textbf{Definition \hypertarget{definition2}{2} (Contractivity).} For an $N$-qubit homogenizer with dynamics governed by (\ref{eq:eq1}, \ref{eq:unitaryoperation}), as in \hyperlink{definition}{Definition 1}, and state evolution given by the iterative transition function (\ref{eq:transition}), the CPTP map $\mathcal{M}$ (\ref{eq:cptp}) is contractive with respect to $S(\mathcal{H}_{\mathcal{R}\times\hat{S}})$ if the inequality (\ref{eq:contraction}) is satisfied for $\forall \rho\in S(\mathcal{H}_{\hat{S}})$ for any pair of which $\rho_{\hat{s}}^{(k)} = \mathcal{M}^{(k)}\rho_{\hat{s}}^{(k-1)}$ holds. That is, for $\forall \rho, \xi \in S(\mathcal{H}_{\mathcal{R}\times\hat{S}})$, $\| \xi^{(k)} - \rho^{(k)}\|_{2} \leq \mathcal{D}^{(k)}$, where the distance function acts like a sequence ($\{\mathcal{D}^{(k)};k\geq 0\}$, where $\mathcal{D}^{(k)}>0$), such that $\lim_{k\rightarrow\infty}\|\mathcal{D}^{(k)}\|_{2}\rightarrow 0$ whenever $\sup_{k\in\mathbb{Z}}\mathcal{D}^{(-k)}(\xi^{(k)}-\rho^{(k)})\rightarrow 0$. A dissipative quantum system, governed by such contractive CPTP map, is asymptotically stable.\\ 

Contraction mapping as a necessary condition for stability plays an essential role in QRC as it endows the reservoir state space with distance function $\mathcal{D}$, induced by a $L_{2}$-norm $\| \cdot \|_{2}$, where for any pair of states at timestep $k$ it reads $\|\varphi^{(k)} -\rho^{(k)}\|_{2}$, and can be trivially extended as $\|\left(\varphi^{(1)}, \dots, \varphi^{(N)}\right) - \left(\rho^{(1)}, \dots, \rho^{(N)}\right)\|_{2} = \text{max}_{k=1,\dots,N}\|\varphi^{(k)} - \rho^{(k)}\|_{2}$, see \cite{contract1, contract2} and \cite{ortega} for a more recent characterization of contractive quantum channels in terms of input-independent steady states. In other words, satisfying \hyperlink{definition2}{Definition 2} suggests the distance between any pair of states, fed into the quantum homogenizer, uniformly vanishes under $\mathcal{M}$. More formally, the condition on the contractivity of (\ref{eq:cptp}) with respect to $S(\mathcal{H}_{\mathcal{R}\times\hat{S}})$ would translate into the statement that for $\forall\varphi,\rho\in S(\mathcal{H}_{\mathcal{R}\times\hat{S}})$, $\| \mathcal{M}\varphi^{(k)} - \mathcal{M}\rho^{(k)} \|_{2} \leq f \|\varphi^{(k-1)}-\rho^{(k-1)}\|_{2}$, where $f$ is a nonlinear function.

\section{\label{sec:level4}Reservoir dynamics}
At each evolution timestep $t=k\tau$ of duration $\tau$, the partial \texttt{$\mathtt{SWAP}$} operator  $\tilde{\mathcal{U}}$ couples the input qubit $\hat{s}$ to one reservoir qubit $\mathcal{R}_{k}$, with the interaction described by the CPTP map \cite{16} 

\begin{equation}
\label{eq:cptp}
\mathcal{M}^{(k)}\left(\rho_{\hat{s}}^{(k)}\right) = \sum_{i}\psi_{i}\, \tilde{\mathcal{U}}\left(\ket{\mathcal{R}_{k}}\otimes\bra{\hat{s}}\right)\tilde{\mathcal{U}}^{\dagger}  
\end{equation}
where $\tilde{\mathcal{U}}$ is given by (\ref{eq:unitaryoperation}), and $\psi$ is a complex amplitude. Here, the linear but nonintegrable reservoir dynamics, induced by the iterative application of (\ref{eq:cptp}), yields the higher-order correlations, where information is spread across exponentially many reservoir degrees of freedom \cite{21}.\footnote{For the purpose of \textit{learning}, the mixing of the higher-order correlations and the linear but nonintegrable dynamics of the CPTP map, is essential.} Suppose the $N$ reservoir qubits are represented as Pauli operator products $\{A_{k}\} = \{\mathbb{1}, X, Y, Z\}^{\otimes N}$ (for $A_{k}A_{j} = \delta_{kj}\mathbb{1}$), where the $k^{\text{th}}$ qubit is given as $\mathcal{R}_{k}=\text{tr}\left(A_{k}\xi\right)$. Then, there exists a state that stores the correlation between $\mathcal{R}_{k}$ and $\mathcal{R}_{j}$, see \cite{7}.

Thus, from a dynamical systems perspective \cite{9}, and assuming $J(t) > 0$, Eq. (\ref{eq:cptp}) acts like an iterative state-transition function with discrete-time nonlinear evolution which, at each timestep, encodes parts of the lower-dimensional input $\hat{s}\in S(\mathcal{H}_{\hat{S}})$ onto a subset of the higher-dimensional reservoir $\mathcal{R}\in S(\mathcal{H}_{\mathcal{R}})$

\begin{equation}
\label{eq:transition}
\mathcal{M}:\mathcal{H}_{\hat{S}}\times\left(\mathcal{H}_{\mathcal{R}}\times\cdots\times \mathcal{H}_{\mathcal{R}}\right)\rightarrow \mathcal{H}_{\mathcal{R}}\times\cdots\times \mathcal{H}_{\mathcal{R}}
\end{equation}
where for a single evolution step, the state update equation reads $\mathcal{M}^{(1)}:\mathcal{H}_{\hat{S}}\times\mathcal{H}_{\mathcal{R}}\rightarrow\mathcal{H}_{\mathcal{R}}$, where $\mathcal{H}_{\hat{S}}$ and $\mathcal{H}_{\mathcal{R}}$ denote, respectively, the input and reservoir state spaces with $\mathcal{H}_{\mathcal{R}} \gg \mathcal{H}_{\hat{S}}$ and $\mathcal{M}\in S(\mathcal{H}_{\mathcal{R}\times\hat{S}})$ \cite{21a}. In that sense, the evolution of the current (echo) state $\Bar{x}_{(k)}$ of an arbitrary input qubit $\hat{s}$ from timestep $(k-1)$ to $k$ is a function of the CPTP map and the input-reservoir coupling strength, such that

\begin{equation}
\label{eq:echostate}
\Bar{x}_{(k)} = f \left(J(t)\hat{s}_{(k)}+\mathcal{M}^{(k)}\Bar{x}_{(k-1)}\right)
\end{equation}
where $J(t)$ acts like an iterative version of the input weight matrix $W_{in}$, found in the feed-forward input layer of a classical RNN \cite{9}, and $f(\cdot)$ is a nonlinear function which contracts all elements towards the steady state $\xi$.

Now, following (\ref{eq:eq1}-\ref{eq:cptp}), we can trivially express the evolution of $\rho^{(k)}_{\hat{s}}$ under $\mathcal{M}$ in a compact form as

\begin{equation}
\label{eq:generalform}
\rho^{(k)}_{\hat{s}} := \left(\prod_{k=1}^{N}\mathcal{M}^{(k)}\right) \rho^{(0)}_{\hat{s}}
\end{equation}
where from (\ref{eq:unitaryoperation}) we know that, explicitly, $\mathcal{M}^{(k)} := p\mathcal{M}^{(k)}_{\mathbb{1}} + (1-p)\mathcal{M}^{(k)}_{\mathtt{S}}$, which is a convex combination that is also a CPTP map, \textit{e.g.} noisy quantum channel \cite{16}. Likewise, upon applying (\ref{eq:cptp}), $\mathcal{M}^{(k)}_{\mathbb{1}}$ and $\mathcal{M}^{(k)}_{\mathtt{S}}$ denote that $\tilde{\mathcal{U}}$ will act, respectively, as either the identity operator or as the $\mathtt{SWAP}$ operator. That is, at each timestep $t=k\tau$ the state $\rho_{\hat{s}}^{(k)}$ will remain unchanged with probability $p$, and with probability $(1-p)$ the states $\rho_{\hat{s}}^{(k)}$ and $\xi^{(k)}$ will be swapped.

Here, $\mathcal{M}$ is \textit{causal}, such that $\forall \bold{u}, \bold{v}\in\mathbb{Z}$ (where $\bold{u}=\{{{u}_{k}}\}^{N}_{k=1}$ and $\bold{v}=\{{{v}_{k}}\}^{N}_{k=1}$), $\mathcal{M}(u_{k})=\mathcal{M}(v_{k})$ $\forall k \in N$, and $u_{\tau}=v_{\tau}$ for $\tau\leq k$ (see, \textit{e.g.}, \cite{8}). Which is to say that the state at time $t$ depends on the evolution up to and including $t$. Besides, given the homogenizer architecture (\ref{eq:eq1}), and from (\ref{eq:transition}), clearly (\ref{eq:generalform}) can be expanded as a pointwise limit due to its recursive nature\footnote{In our setting, the CPTP map is kept input-independent, \textit{i.e.} time-invariant in the course of the evolution.}

\begin{equation}
\label{eq:explicitform}
\rho^{(k)}_{\hat{s}} = \mathcal{M}^{(k)}\circ\mathcal{M}^{(k-1)}\circ\cdots\circ\mathcal{M}^{(1)}\left(\rho^{(0)}_{\hat{s}} \right)
\end{equation}
which is both \textit{well-defined} and independent of $\rho^{(0)}_{\hat{s}}$. Meaning, suppose $\mathcal{M}:S_{\text{in}}\rightarrow S_{\sigma}$, where $S_{\text{in}}$ and $S_{\sigma}\subseteq\mathcal{R}$ denote the set of operators for, respectively, the logical input state and a subset of the reservoir in state $\sigma$. Then, for all $t\in \mathbb{Z_{+}}$, and for all sets of real-valued sequences $\bold{u}, \bold{v}\in\mathbb{Z}$, $\sup_{t\in\mathbb{Z}}|\mathcal{M}_{\bold{u}}-\mathcal{M}_{\bold{v}}|_{t} := |(\mathcal{M}_{u}\mathcal{M}_{u-1}\dots \rho_{\hat{s}}^{(-N)}) - (\mathcal{M}_{v}\mathcal{M}_{v-1}\dots \rho_{\hat{s}}^{(-N)})| < \delta$, where $\delta>0$ and $\sup_{t\in\mathbb{Z}}|\cdot|_{t} :=  \| \cdot \|_{\infty}$ \cite{rudin}. Here, $\bold{u}$ and $\bold{v}$ are the limit of the above sequence, with scalar product given by $\langle\bold{u},\bold{v}\rangle_{\mathcal{M}^{(N)}} := \lim_{k\rightarrow\infty}\langle u_{k},v_{k} \rangle$, where $u_{k}\rightarrow v_{k}$. If we assume that the set of operators $S(\mathcal{H}_{\mathcal{R}})$ forms a complete metric space, induced by the $L_{2}$-norm \cite{metric}, and given that any input sequence converges to a point in $S(\mathcal{H}_{\mathcal{R}})$, then $\{\mathcal{M}\}^{\otimes N}$ forms a Cauchy sequence \cite{8}.

Besides, as a consequence of the Banach theorem \cite{22,23}, the iterative application of (\ref{eq:cptp}) yields \textit{convergence} to a fixed point $\mathcal{M}^{(N)}(\rho)\rightarrow\xi$. Where in our setting (as shown in \cite{16}), the convergence is towards the only fixed point in $S(\mathcal{H}_{\mathcal{R}})$ for which $\mathcal{M}(\xi)=\xi$, namely the desired steady state $\xi^{(0)}\in S(\mathcal{H}_{\mathcal{R}})$.\footnote{Here, $\xi^{(0)}\in S(\mathcal{H}_{\mathcal{R}})$ can be viewed as the natural analog of the \textit{unique density operator}, found in the mixing property \cite{mixing}, where as we showed in \cite{16}, the convergence is done via a noisy quantum channel, given by an input-independent CPTP map. Therefore, further supporting the idea that convergence is an extension of the mixing property.} This implies that \cite{mixing}

\begin{equation}
\lim_{N\rightarrow\infty} \| \Sigma_{N}^{(\mathcal{M})}\left(\rho_{\hat{s}}^{(0)}\right) - \xi^{(0)} \|_{2} = 0
\end{equation}
where 

\begin{equation}
\Sigma_{N}^{(\mathcal{M})}\left(\rho_{\hat{s}}^{(0)}\right) := \frac{1}{N}\sum_{k=0}^{N}\mathcal{M}^{(k)}\left(\rho_{\hat{s}}^{(0)}\right)
\end{equation}
denotes the average state after $N$ repetitive applications of $\mathcal{M}$ to $\rho_{\hat{s}}^{(0)}$.

From \hyperlink{definition2}{Definition 2}, let $\mathcal{D}(\rho, \xi)$ be a differentiable distance function. Then, the CPTP map (\ref{eq:cptp}) is \textit{contractive} if $\forall \rho, \xi \in S(\mathcal{H}_{\mathcal{R}\times\hat{S}})$, $\mathcal{D}\left(\mathcal{M}(\rho), \mathcal{M}(\xi)\right)\leq k\mathcal{D}(\rho,\xi)$, where $0\leq k < 1$, and if the following inequality is satisfied  

\begin{equation}
\label{eq:contraction}
\| \xi^{(k)} - \rho_{\hat{s}}^{(k)} \|_{2} \leq \| \xi^{(k-1)} - \rho_{\hat{s}}^{(k-1)} \|_{2}
\end{equation}
where the state distance is induced by the $L_{2}$-norm.

To prove the satisfiability, let $\|\mathcal{D}^{(k)}\|_{2} \equiv \| \xi^{(k)} - \rho_{\hat{s}}^{(k)} \|_{2}$, and assume for simplicity that $k=1$. Then, from (\ref{eq:explicitform}) and for $\forall \rho, \xi \in S(\mathcal{H}_{\mathcal{R}\times\hat{S}})$

\begin{equation}
\begin{split}
\|\mathcal{D}^{(k)}\|_{2} & = \|\mathcal{M}^{(k)} \left(\xi^{(0)} - \rho_{\hat{s}}^{(0)}\right) \|_{2} \\
& = \| \mathcal{M}^{(k)}\left(\xi^{(0)}\right) - \mathcal{M}^{(k)}\left(\rho_{\hat{s}}^{(0)}\right) \|_{2} \\
& \leq \| \mathcal{M}^{(1)} \|_{2} \, \cdot \|\mathcal{D}^{(k-1)}\|_{2} \\
& = \mathcal{M}^{(1)} \|\mathcal{D}^{(k-1)}\|_{2} 
\end{split}
\end{equation}
Clearly, by running the homogenizer long enough (having a large number of input-reservoir interactions), we get $\lim_{k\rightarrow\infty}\|\mathcal{D}^{(k)}\|_{2}\rightarrow 0$, where the relation between state fidelity and the reservoir size was examined numerically in \cite{anna}. Physically, it is the natural convergence of the homogenizer to a desired steady state (see, \textit{e.g.}, \cite{16}) that guarantees the distance of an arbitrary logical state with respect to $\xi^{(0)}\in S(\mathcal{H}_{\mathcal{R}})$ at timestep $k$ will be less than the distance at timestep $(k-1)$ as it contracts all elements of the state space \cite{contract1, contract2}. Therefore, we see that the CPTP map $\mathcal{M}$ is \textit{contractive} (\hyperlink{definition2}{Definition 2}) which is sufficient to prove that the homogenizer has \textit{convergent} (asymptotically stable) dynamics, as in \hyperlink{definition}{Definition 1} \cite{40}. Namely, it is evident that in this setting $\lim_{N\rightarrow\infty} \parallel\prod^{N}_{k=1}\mathcal{M}^{(k)}\left(\rho_{\hat{s}}^{(0)}\right)\parallel_{2} = \xi^{(0)}$.

This highlights the role of the homogenizer dynamics for processing temporal input data. Its ability to engineer readily-available steady states for arbitrary inputs \cite{16, 17, 17a} is crucial for avoiding convergence to a maximally-mixed state $\rho^{(N)}_{\hat{s}}=\mathbb{1}^{\otimes N} / 2^{N}$ \cite{mixed1, mixed2}, which would otherwise lead to asymptotically vanishing Volterra kernels (poor information retention), even for finite-size reservoirs. Interestingly, this was also recognized in \cite{16}, and independently in \cite{kernels}, to yield non-trivial \textit{quantum} Volterra kernels (which describe the system's \textit{memory}, or dependence on past timesteps), that are robust to dissipative dynamics, and the noisy environment of current quantum hardware.

\begin{figure}[ht]
\centering
\includegraphics[width=\linewidth]{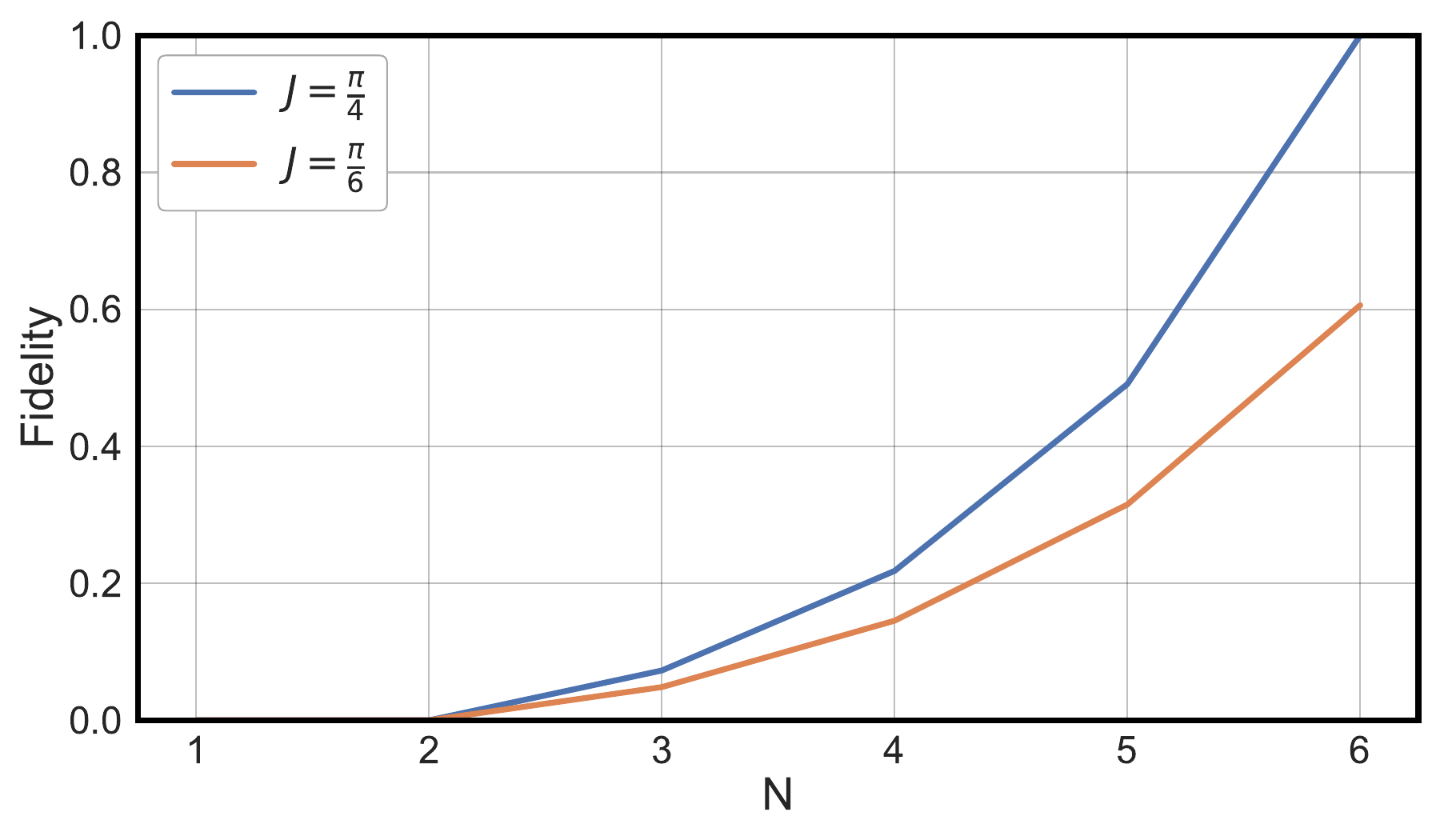}
\caption{Numerical simulation illustrating the convergence rate (\textit{i.e.} state transformation, measured by the fidelity, as in \cite{vlatko2, fidelity}) of an input qubit that sequentially interacts with $N = 5$ identically-prepared reservoir qubits under the described homogenizer dynamics for two different coupling values, $J=\pi/4$ (blue line) and $J=\pi/6$ (orange line). Note that, as highlighted in \cite{vlatko2}, the closer $J$ is to $\pi/2$, the higher the probability is that $\tilde{\mathcal{U}}$ (as in Eq. (\ref{eq:unitaryoperation})) will act as the standard \texttt{SWAP} operation. Evidently, as the coupling strength increases, the homogenizer is more effective; which introduces a symmetry between the roles of the input system and the reservoir.}
\label{fig:figure2}
\end{figure}

On the other hand, as discussed in \cite{16,17}, and also showed here, the dominant term in the dissipation dynamics of the homogenizer is the time-dependent coupling $J(t)$. The behavior of the homogenizer under the influence of $J(t)$ could be understood as follows, see Fig. \ref{fig:figure2}. If the coupling between the input and the $k^{\text{th}}$ reservoir qubit is vanishing, $J(t)\sim 0$, then $\tilde{\mathcal{U}}\equiv\mathbb{1}$, and no information will be encoded in that qubit. Hence, the distance between the input and reservoir density matrices (\ref{eq:contraction}) remains invariant. From a practical point of view, one could try to fix $J(t)$ to a small non-zero value and simply increase the size of the reservoir. Although this would improve the memory \cite{24}, it would have little effect on the efficiency of the convergent mapping as: (i) $\tilde{\mathcal{U}}$ would act as $\mathbb{1}$ with higher probability, and (ii) longer run times complicate the scalability due to decoherence. 

On the upside, however, here the homogenizer remains relatively unchanged by the 2-qubit interactions and could be reused for several inputs without the need for reinitialization after each one; meaning it obtains continuing temporal memory for longer data streams. If we increase the strength of $J(t)$, on the other hand, the convergence dynamics becomes much more efficient and the relaxation time between consecutive input-reservoir interactions sharply decreases. Thus, this allows for random mapping of the information, encoded in the input density matrix, in a nonlinear form. While preferable for rapid and efficient homogenization, the strong coupling ruins reusability and has negative effects on the long-term memory of the reservoir.

Evidently, in this new QRC implementation the iterative application of the CPTP map is similar to the action of stacked recurrent hidden layers in deep networks, where the operation of $\tilde{\mathcal{U}}$ each timestep is akin to that of a reservoir layer described by converging discrete-time dynamics, while $J(t)$ (as in Eq. (\ref{eq:echostate})) acts like a discrete-time version of the random input weight matrix, found in RNN models, whose role is to map the input to the reservoir \cite{25}.

Experimentally, there have been several realizations of this protocol on different platforms, showing consistency with theoretical predictions. In \cite{vlatko2} both the performance and robustness to deterioration of a 3-qubit homogenization machine were demonstrated using single-photon qubits, produced by a low-noise heralded source, where the partial \texttt{SWAP} was implemented via a series of fiber beam splitters. While in \cite{17}, the homogenizer was realized on a liquid-state NMR quantum processor, given by four $^{13}\text{C}$ nuclei in a sample of fully $^{13}\text{C}$-labelled crotonic acid dissolved in deuterated acetone. Implementations using electron spins are also possible, particularly in the context of exchange-interaction universal quantum computing \cite{levy}; here the 2-qubit operations (modulated by the system-reservoir coupling $J$) can be controlled by electro-magnetic fields, where the electron charge controls the strength of $J$. Remarkably, the model shows immunity from collective decoherence and significant reduction in hardware resources. Overall, future implementations of the homogenizer on various quantum technologies (\textit{e.g.} spin-based architectures \cite{levy}) can shed light on possible optimizations of existing quantum protocols, \textit{e.g.} quantum steady state preparation, stabilization, and error mitigation; as well as motivate the development of novel quantum thermodynamics-inspired devices.

\section{\label{sec:level5}Conclusions}
Lately, QRC was recognized as a promising platform for solving temporal ML problems. It is an appealing alternative to quantum gate-based models, as the computations are outsourced to the reservoir without need for parameter fine-tuning. Nonetheless, engineering such quantum systems with available hardware is challenging since their efficiency is highly dependent on the reservoir dynamics being \textit{asymptotically stable} and \textit{contractive}. Besides, certain robustness to noise, and ability to produce non-trivial steady states for arbitrary inputs \cite{16} was recently recognized as a necessary condition for generating non-vanishing \textit{quantum} Volterra kernels \cite{kernels} and memory retention.

In this work, we bridged the gap between QRC and dissipative collision models, and extended the quantum homogenizer as an alternative reservoir-based platform for temporal ML. Studying its dynamics, induced by the iterative discrete-time application of a unitary bipartite operator, we demonstrated that the time evolution of an input density matrix, injected into the reservoir, is given by a CPTP map which is \textit{contractive}, thus satisfying \hyperlink{definition2}{Definition 2}. More importantly, this is a sufficient requirement to prove that the reservoir dynamics is also \textit{asymptotically stable}, as in \hyperlink{definition}{Definition 1}. By proving the homogenizer satisfies those necessary and sufficient conditions, and combined with its natural ability to produce target steady states (which endows it with \textit{persistent memory} in the context of \cite{kernels}), we establish it as a viable quantum hardware platform. In future work we will evaluate the computational capabilities and memory capacity of the homogenizer on benchmark tasks such as NARMA (nonlinear autoregressive moving-average).

\begin{acknowledgments}
V.V. thanks the Oxford Martin School, the John Templeton Foundation, the Gordon and Betty Moore Foundation, and the EPSRC (UK). 
\end{acknowledgments}

\end{document}